\documentclass[preprints,article,accept,moreauthors,10pt,a4paper]{mdpi}

\firstpage{1}
\pubvolume{xx}
\issuenum{1}
\articlenumber{1}
\pubyear{2018}
\copyrightyear{2018}
\externaleditor{Academic Editor: name}
\history{Received: date; Accepted: date; Published: date}

\usepackage{amssymb}
\usepackage{hyperref}
\usepackage[hyphenbreaks]{breakurl}
    \newcommand\beq{\begin{equation}}
\newcommand\eeq{\end{equation}}
\newcommand\beqa{\begin{eqnarray}}
\newcommand\eeqa{\end{eqnarray}}
\newcommand{\nn}{\nonumber\\}
\def\bal#1\eal{\begin{align}#1\end{align}}

\newcommand{\ex}{{\text{ex}}}
\newcommand{\id}{{\text{id}}}
\newcommand{\dd}{{\text{d}}}
\newcommand{\Ds}{\widetilde{s}}
\newcommand{\di}{d}
\newcommand{\con}{{\text{c}}}
\newcommand{\vir}{{\text{vir}}}
\newcommand{\comp}{{\text{comp}}}
\newcommand{\eeta}{\phi}
\newcommand{\on}{\text{1D}}
\newcommand{\thr}{\text{3D}}
\newcommand{\eff}{\text{eff}}

\Title{Residual Multiparticle Entropy for a Fractal Fluid of Hard Spheres}

\newcommand{\orcidauthorA}{0000-0002-9564-5180} % Add \orcidA{} behind the author's name
\newcommand{\orcidauthorB}{0000-0002-5970-9001} % Add \orcidB{} behind the author's name
\newcommand{\orcidauthorC}{0000-0002-4851-5105} % Add \orcidC{} behind the author's name

\Author{Andr\'es Santos $^{1,}$*\orcidA{}, Franz Saija $^{2}$\orcidB{} and Paolo V. Giaquinta $^{3}$\orcidC{}}

\AuthorNames{Andr\'es Santos, Franz Saija  and Paolo V. Giaquinta}

\address{%
$^{1}$ \quad Departamento de F\'{\i}sica and Instituto de Computaci\'on Cient\'ifica Avanzada (ICCAEx), Universidad de Extremadura,
E-06006 Badajoz, Spain; andres@unex.es\\
$^{2}$ \quad CNR-IPCF, Viale F. Stagno d'Alcontres, 37-98158 Messina, Italy; saija@ipcf.cnr.it\\
$^{3}$ \quad  Universit\`a degli Studi di Messina, Dipartimento di Scienze Matematiche e Informatiche, Scienze Fisiche e Scienze della Terra, Contrada Papardo, 98166 Messina, Italy; paolo.giaquinta@unime.it}

\corres{Correspondence: andres@unex.es; Tel.: +34-924-289-651}

\abstract{
The residual multiparticle entropy (RMPE) of a fluid is defined as the difference, $\Delta s$, between the excess entropy per particle (relative to an ideal gas with the same temperature and density), $s_\text{ex}$, and the pair-correlation contribution, $s_2$. Thus, the RMPE represents the net contribution to $s_\text{ex}$ due to spatial correlations involving three, four, or more particles. A heuristic ``ordering'' criterion identifies the vanishing of the RMPE as an underlying signature of an impending structural or thermodynamic transition of the system from a less ordered to a more spatially organized condition (freezing is a typical example).  Regardless of this, the knowledge of the RMPE is important to assess the impact of non-pair multiparticle correlations on the entropy of the fluid. Recently, an accurate and simple proposal for the thermodynamic and structural properties of a hard-sphere fluid in fractional dimension $1<d<3$ has been proposed [Santos, A.; L\'opez de Haro, M. \emph{Phys. Rev. E} \textbf{2016}, \emph{93}, 062126]. The aim of this work is to use this approach to evaluate the RMPE as a function of both $d$ and the packing fraction $\phi$. It is observed that, for any given dimensionality $d$, the RMPE takes negative values for small densities, reaches a negative minimum $\Delta s_{\text{min}}$ at a packing fraction $\phi_{\text{min}}$, and then rapidly increases, becoming positive beyond a certain packing fraction $\phi_0$. Interestingly, while both $\phi_{\text{min}}$ and $\phi_0$  monotonically decrease as dimensionality increases, the value of $\Delta s_{\text{min}}$ exhibits a nonmonotonic behavior, reaching an absolute minimum at a fractional dimensionality $d\simeq 2.38$.
A plot of the scaled RMPE $\Delta s/|\Delta s_{\text{min}}|$ shows a quasiuniversal behavior in the region $-0.14\lesssim\phi-\phi_0\lesssim 0.02$.
}


\begin{document}
%%%%%%%%%%%%%%%%%%%%%%%%%%%%%%%%%%%%%%%%%%
%% Only for the journal Gels: Please place the Experimental Section after the Conclusions

%%%%%%%%%%%%%%%%%%%%%%%%%%%%%%%%%%%%%%%%%%
%\setcounter{section}{-1} %% Remove this when starting to work on the template.
%\section{How to Use this Template}
%The template details the sections that can be used in a manuscript. Note that the order and names of article sections may differ from the requirements of the journal (e.g. the positioning of the Materials and Methods section). Please check the instructions for authors page of the journal to verify the correct order and names. For any questions, please contact the editorial office of the journal or support@mdpi.com. For LaTeX related questions please contact Janine Daum at latex-support@mdpi.com.
%The order of the section titles is: Introduction, Materials and Methods, Results, Discussion, Conclusions for these journals: aerospace,algorithms,antibodies,antioxidants,atmosphere,axioms,biomedicines,carbon,crystals,designs,diagnostics,environments,fermentation,fluids,forests,fractalfract,informatics,information,inventions,jfmk,jrfm,lubricants,neonatalscreening,neuroglia,particles,pharmaceutics,polymers,processes,technologies,viruses,vision

\section{Introduction}
\label{sec1}
The properties of liquids are of great interest in many science and engineering areas. Aside from ordinary three-dimensional systems, many interesting phenomena do also occur in restricted one- or two-dimensional geometries, under the effect of spatial confinement. Actually, there are also cases where the configuration space exhibits, at suitable length scales, non-integer dimensions. Indeed, many aggregation and growth processes can be described quite well by resorting to the concepts of fractal geometry. This is the case, for example, of liquids confined in porous media or of assemblies of small particles forming low-density clusters and networks \cite{WC92,KCK09,KMS11,SSAHD13}.

Heinen \textit{et al.} \cite{HSBL15} generalized this issue by introducing fractal particles in a fractal configuration space. In their framework the particles composing the liquid are fractal as is the configuration space in which such objects move. Santos and L\'opez de Haro \cite{SH16} have further developed reliable heuristic interpolations for the equation of state and radial distribution function of hard-core fluids in fractal dimensions between one and three dimensions. Taking advantage of their work, we intend to study in this paper some thermostatistical properties of such fractal systems in the theoretical framework provided by the multiparticle correlation expansion of the excess entropy,
\beq
\label{1}
s_\ex(\rho,\beta)=s(\rho,\beta)-s_\id(\rho,\beta),
\eeq
where $\rho$ is the number density, $\beta=1/k_BT$ is the inverse temperature, $s(\rho,\beta)$ is the entropy per particle (in units of the Boltzmann constant $k_B$), and
\beq
\label{s_id}
s_\id(\rho,\beta)=\frac{d+2}{2}-\ln\left[\rho\left(\frac{h^2\beta}{2\pi m}\right)^{d/2}\right]
\eeq
is the ideal-gas entropy; in Eq.\ \eqref{s_id},  $d$ is the spatial dimensionality of the system, $h$ is Planck's constant, and $m$ is the mass of a particle.

As is well known, the excess entropy can be expressed as an infinite sum of contributions associated with spatially integrated density correlations of increasing order \cite{Nettleton1958,Baranyai1989}. In the absence of external fields, the leading and quantitatively dominant term of the series is the so-called ``pair entropy'',
\beq
s_2(\rho,\beta)=-\frac{\rho}{2}\int \dd\mathbf{r}\,\left[g(r;\rho,\beta)\ln g(r;\rho,\beta)-g(r;\rho,\beta)+1\right],
\label{2a}
\eeq
whose calculation solely requires the knowledge of the pair distribution function of the fluid, $g(r;\rho,\beta)$.
An integrated measure of the importance of more-than-two-particle density correlations in the overall entropy balance is given by the so-called ``residual multiparticle entropy'' (RMPE) \cite{GG92}:
\beq
\label{rmpe}
\Delta s(\rho,\beta)=s_\text{ex}(\rho,\beta)-s_2(\rho,\beta).
\eeq
It is important to note that, at variance with $s_\ex$ and $s_2$, which are both negative definite quantities, $\Delta s$ may be either negative or positive. As originally shown by Giaquinta and Giunta for hard spheres in three dimensions \cite{GG92}, the sign of this latter quantity does actually depend on the thermodynamic state of the fluid. In fact, the RMPE of a hard-sphere fluid is negative at low densities, thus contributing to a global reduction of the phase space available to the system as compared to the corresponding ideal gas. However, the RMPE undergoes a sharp crossover from negative to positive values at a value of the packing fraction which substantially overlaps with the thermodynamic freezing threshold of the hard-sphere fluid. Such a behavior suggests that at high enough densities multiparticle correlations play an opposite role with respect to that exhibited in a low packing regime in that they temper the decrease of the excess entropy that is largely driven by the pair entropy. The change of sign exhibited by the RMPE is a background indication, intrinsic to the fluid phase, that particles, forced by more and more demanding packing constraints, start exploring, on a local scale, a different structural condition. This process is made possible by an increasing degree of cooperativity, that is signalled by the positive values attained by $\Delta s$, which gradually leads to a more efficacious spatial organization and ultimately triggers the crystalline ordering of the system on a global scale.

A similar indication is also present in the RMPE of hard rods in one dimension \cite{G08}. In this model system, notwithstanding the absence of a fluid-to-solid transition, one can actually observe the emergence of a solid-like arrangement at high enough densities: tightly-packed particles spontaneously confine themselves within equipartitioned intervals whose average length is equal to the the total length per particle, even if the onset of a proper entropy-driven phase transition is frustrated by topological reasons. Again, even in this ``pathological'' case, the vanishing of the RMPE shows up as an underlying signature of a structural change which eventually leads to a more ordered arrangement.

The relation between the zero-RMPE threshold and the freezing transition of hard spheres apparently weakens in four and five dimensions \cite{KSET08}, where lower bounds of the entropy threshold significantly overshoot the currently available computer estimates of the freezing density \cite{KSET08,KSET09}. On the other side, a close correspondence between the sign crossover of the RMPE and structural or thermodynamical transition thresholds has been highlighted in both two and three dimensions on a variety of model systems for different macroscopic ordering phenomena other than freezing \cite{G09}, including fluid demixing \cite{Saija1998}, the emergence of mesophases in liquid crystals \cite{Costa2002}, the formation of a hydrogen-bonded network in water \cite{Saija2003}, or, more recently, the onset of glassy dynamics \cite{BNSMB2017}.

If hard-core systems in fractal geometries exhibit a sort of disorder-to-order transition, it seems plausible that such a transition is signaled by a change of sign of $\Delta s$.
Taking all of this into account, it is desirable to study the RMPE of  hard-core fractal fluids, and this is the main goal of this paper.
It is organized as follows. The theoretical approach of Ref.\ \cite{SH16} is described and applied to the evaluation of the RMPE in Sec.\ \ref{sec2}.
The results are presented and discussed in Sec.\ \ref{sec3}. Finally, the main conclusions of the work are recapped in Sec.\ \ref{sec4}.

%%%%%%%%%%%%%%%%%%%%%%%%%%%%%%%%%%%%%%%%%%

\section{Methods}
\label{sec2}
\subsection{General relations}
In principle, the knowledge of the pair distribution function, $g(r;\rho,\beta)$, allows one to determine the pair entropy from Eq. \eqref{2a}. This is equivalent to
\beq
s_2(\rho,\beta)=\frac{1}{2}\left[\chi_T(\rho,\beta)-1\right]+\Ds_2(\rho,\beta),
\label{2b}
\eeq
where
\beq
\chi_T(\rho,\beta)=1+\rho \int \dd\mathbf{r}\,\left[g(r;\rho,\beta)-1\right]
\label{chiT}
\eeq
is the isothermal susceptibility and we have called
\beq
\Ds_2(\rho,\beta)=-\frac{\rho}{2}\int \dd\mathbf{r}\,g(r;\rho,\beta)\ln g(r;\rho,\beta).
\label{2}
\eeq
Thus, Eq. \eqref{rmpe} can be rewritten as
\beq
\Delta s(\rho,\beta)=s_\text{ex}(\rho,\beta)-\frac{1}{2}\left[\chi_T(\rho,\beta)-1\right]-\Ds_2(\rho,\beta).
\label{Delta_s}
\eeq

\noindent Equations \eqref{2b}--\eqref{Delta_s} hold regardless of whether the total potential energy $U(\mathbf{r}_1, \mathbf{r}_2, \mathbf{r}_3,\ldots)$ is pairwise additive or not. On the other hand, if $U$ is pairwise additive, the knowledge of $g(r;\rho,\beta)$ yields, apart from $s_2(\rho,\beta)$, the thermodynamic quantities of the system via the so-called thermodynamic routes \cite{S16}. In particular, the virial route is
\bal
\label{ZZ}
Z(\rho,\beta)\equiv\frac{\beta p(\rho,\beta)}{\rho}=&1-\frac{\rho \beta}{2d}\int \dd\mathbf{r}\,r\frac{\dd u(r)}{\dd r}g(r;\rho,\beta)\nn
=&1+\frac{\rho}{2d}\int \dd\mathbf{r}\,r\frac{\dd e^{-\beta u(r)}}{\dd r}y(r;\rho,\beta),
\eal
where $p$ is the pressure, $Z$ is the compressibility factor,  $u(r)$ is the pair interaction potential, and $y(r;\rho,\beta)\equiv e^{\beta u(r)}g(r;\rho,\beta)$ is the so-called cavity function. Next, the excess Helmholtz free energy per particle, $a_\ex$, and the excess entropy per particle, $s_\ex$, can be obtained by standard thermodynamic relations as
\beq
\label{betaa_ex}
\beta a_\ex(\rho,\beta)=\int_0^1\dd t\frac{Z(\rho t,\beta)-1}{t},\quad {s_\ex(\rho,\beta)}=\beta\frac{\partial \beta a_\ex(\rho,\beta)}{\partial\beta}-\beta a_\ex(\rho,\beta).
\eeq
Combining Eqs.\ \eqref{ZZ} and \eqref{betaa_ex}, we obtain
\beq
\label{s_ex}
s_\ex(\rho,\beta)=\frac{\rho}{2d}\left(\beta\frac{\partial}{\partial\beta}-1\right)
\int \dd\mathbf{r}\,r\frac{\dd e^{-\beta u(r)}}{\dd r}\int_0^1\dd t\,y(r;\rho t,\beta).
\eeq

To sum up, assuming the pair distribution function $g(r;\rho,\beta)$ for a $d$-dimensional fluid of particles interacting via an interaction potential $u(r)$ is known, it is possible to determine the excess entropy [see Eq. \eqref{1}], the pair entropy [see Eq.\ \eqref{2a}], and hence the RMPE $\Delta s$. Note that, while $s_2$ only requires $g(r)$ at the state point $(\rho,\beta)$ of interest, $s_\ex$ requires the knowledge of $g(r)$ at all densities smaller than $\rho$ and at inverse temperatures in the neighborhood of $\beta$.

A remark is now in order. The isothermal susceptibility $\chi_T(\rho,\beta)$ can be obtained directly from $g(r;\rho,\beta)$ via Eq.\ \eqref{chiT} or indirectly via Eq.\ \eqref{ZZ} and the thermodynamic relation
\beq
\chi_T^{-1}(\rho,\beta)=\frac{\partial \rho Z(\rho,\beta)}{\partial\rho}.
\label{4}
\eeq
If the correlation function $g(r;\rho,\beta)$ is determined from an approximate theory, the compressibility route \eqref{chiT} and the virial route given by Eqs.\ \eqref{ZZ} and \eqref{4} yield, in general, different results.

\subsection{Fractal hard spheres}
Now we particularize to hard-sphere fluids in $d$ dimensions. The interaction potential is simply given by
\beq
u(r)=\begin{cases}
  \infty,& r<\sigma,\\
  0, & r>\sigma,
\end{cases}
\eeq
where $\sigma$ is the diameter of a sphere. In this case, the pair distribution function $g(r;\eeta)$ is independent of temperature and the density can be characterized by the packing fraction
\beq
\label{eeta}
\eeta\equiv \frac{(\pi/4)^{d/2}}{\Gamma(1+d/2)}\rho\sigma^d.
\eeq
Taking into account that $\frac{\dd}{\dd r} e^{-\beta u(r)}=\delta(r-\sigma)$, Eqs.\ \eqref{ZZ} and \eqref{s_ex} become
\beq
Z(\eeta)=1+2^{d-1}\eeta g_\con(\eeta),
\label{Z}
\eeq
\beq
s_\ex(\eeta)=-\beta a_\ex(\eeta)=2^{d-1}\eeta\int_0^1\dd t\, g_\con(\eeta t),
\label{5}
\eeq
where $g_\con(\eeta)=g(\sigma^+;\eeta)=y(\sigma;\eeta)$ is the \emph{contact} value of the pair distribution function. Also, Eq.\ \eqref{2} can be written as
\beq
\Ds_2(\eeta)=-d 2^{d-1}\eeta \int_0^\infty \dd r\, r^{d-1} g(r;\eeta)\ln g(r;\eeta).
\label{8b}
\eeq

\noindent In Eqs.\ \eqref{eeta}--\eqref{8b} it is implicitly assumed that $d$ is an integer dimensionality. However, in a pioneering work \cite{HSBL15} Heinen \emph{et al.} introduced the concept of classical liquids in fractal dimension and performed Monte Carlo (MC) simulations of fractal ``spheres'' in a fractal configuration space, both with the same noninteger dimension. Such a generic model of fractal liquids can describe, for instance, microphase separated binary liquids in porous media and highly branched liquid droplets confined to a fractal polymer backbone in a gel. For a discussion on the use of two-point correlation functions in fractal spaces, see Ref.\ \cite{LS91}.

It seems worthwhile extending Eqs.\ \eqref{eeta}--\eqref{8b} to a noninteger dimension $d$ and studying the behavior of the RMPE $\Delta s$ as a function of both $\eeta$ and $d$. To this end, an approximate theory providing the pair distribution function $g(r;\eeta)$ for noninteger $d$ is needed. In Ref.\ \cite{HSBL15}, Heinen \emph{et al.} solved numerically the Ornstein--Zernike relation \cite{BH76} by means of the Percus--Yevick (PY) closure \cite{PY58}. However, since one needs to carry out an integration in Eq.\ \eqref{8b} over all distances, an analytic approximation for $g(r;\eeta)$ seems highly desirable.

In Ref.\ \cite{SH16} a  simple analytic approach was proposed for the thermodynamic and structural properties of the fractal hard-sphere fluid. Comparison with MC simulation results for $d=1.67659$ showed results comparable to or even better than those obtained from the numerical solution of the PY integral equation. In this approach the contact value of the pair distribution function is approximated by
\beq
g_\con(\eeta)=\frac{1-k_\di\eeta}{(1-\eeta)^2},
\label{g1-dD}
\eeq
with
\beq
\label{g1-dD_a2}
k_\di=\frac{(5-d)(2-d)}{4}+(3-d)(d-1)k_2,\quad k_2=\frac{2\sqrt{3}}{\pi}-\frac{2}{3}\simeq 0.436.
\eeq
When particularized to $d=1$, $2$, and $3$, Eq.\ \eqref{g1-dD}  gives the exact \cite{S16}, the Henderson \cite{H75}, and the  PY \cite{W63,T63} results, respectively.
Insertion into Eq.\ \eqref{Z} gives the compressibility factor $Z(\eeta)$ and, by application of Eq.\ \eqref{4}, the isothermal susceptibility as
\beq
\chi_T(\eeta)=\left[1+2^{d-1}\eeta\frac{2-k_\di\eeta(3-\eeta)}{(1-\eeta)^3}\right]^{-1}.
\label{7}
\eeq
Analogously, Eq.\ \eqref{5} yields
\beq
s_\ex(\eeta)=-2^{d-1}\left[\frac{(1-k_\di)\eeta}{1-\eeta}-k_\di\ln(1-\eeta)\right].
\label{6b}
\eeq

\noindent Thus, in order to complete the determination of $\Delta s$ from Eq.\ \eqref{Delta_s}, only $\Ds_2$ remains. It requires the knowledge of the full pair distribution function [see Eq.\ \eqref{8b}].
In the approximation of Ref.\ \cite{SH16}, $g(r;\eeta)$ is given by the simple interpolation formula
\beq
g(r;\eeta)=\alpha(\eeta) g_\on\left(r;\eeta_\on^\eff(\eeta)\right)+[1-\alpha(\eeta)]g_\thr\left(r;\eeta_\thr^\eff(\eeta)\right),
\label{6}
\eeq
where $g_\on(r;\eeta)$ and $g_\thr(r;\eeta)$ are the exact and PY functions for $d=1$ and $3$, respectively,
\beq
\label{8}
\eeta_\on^\eff(\eeta)=\frac{g_\con(\eeta)-1}{g_\con(\eeta)},\quad
\eeta_\thr^\eff(\eeta)=\frac{1+4g_\con(\eeta)-\sqrt{1+24g_\con(\eeta)}}{4 g_\con(\eeta)}
\eeq
are effective packing fractions, and
\beq
\alpha(\eeta)=\frac{H(\eeta)-H_\thr\left(\eeta_\thr^\eff(\eeta)\right)}{H_\on\left(\eeta_\on^\eff(\eeta)\right)-H_\thr\left(\eeta_\thr^\eff(\eeta)\right)}
\label{10}
\eeq
is the mixing parameter.
In Eq.\ \eqref{10},
\beq
\label{HH}
H(\eeta)=\frac{\frac{1}{2}-A_\di\eeta+C_\di\eeta^2}{1+(d-1)\eeta\left[1+(3-d)(1-2k_2)(3-\eeta)\eeta\right]},
\eeq
with
\beq
A_\di=\frac{(2-d)(63-23d)}{60}+\frac{3(d-1)(3-d)}{4}k_2,\quad
C_\di=\frac{(2-d)(8-3d)}{20}+\frac{(d-1)(3-d)}{4}k_2.
\eeq
Of course, $H_\on(\eeta)$ and $H_\thr(\eeta)$ are obtained from Eq.\ \eqref{HH} by setting $d=1$ and $d=3$, respectively.

Summing up, the proposal of Ref.\ \cite{SH16} for noninteger $d$  is defined by Eqs.\ \eqref{6}--\eqref{10}, with $g_\con(\eeta)$ and $H(\eeta)$ being given by Eqs.\ \eqref{g1-dD} and \eqref{HH}, respectively.
By construction, this approximation  reduces to the exact and PY results in the limits $d\to 1$ and $d\to 3$, respectively. Moreover, it is consistent  (via both the virial and compressibility routes) with Henderson's equation of state \cite{H75} in the limit $d\to 2$. The corresponding  isothermal susceptibility and excess free energy are given by Eqs.\ \eqref{7} and \eqref{6b}.
Finally, $\Delta s(\eeta)$ can be obtained from Eq.\ \eqref{Delta_s} by evaluating $\Ds_2(\eeta)$ from Eq.\ \eqref{8b} numerically. To that end, and in order to avoid finite-size effects, it is convenient to split the integration range $0< r<\infty$  into $0<r<R$ and $R<r<\infty$, with $R=10\sigma$. In the first integral  the analytically known function $g(r;\eeta)$ is used, while in the second integral $g(r;\eeta)$ is replaced by its asymptotic form \cite{SH16}.
\section{Results and Discussion}
\label{sec3}
\begin{figure}[H]
\centering
\includegraphics[width=.45\columnwidth]{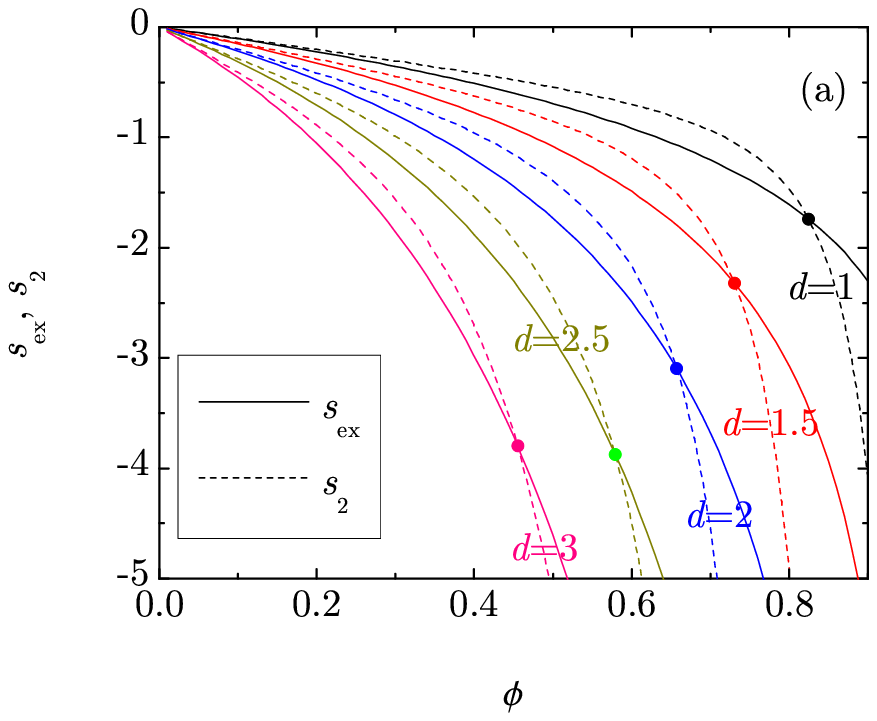}\hspace{1cm}\includegraphics[width=.45\columnwidth]{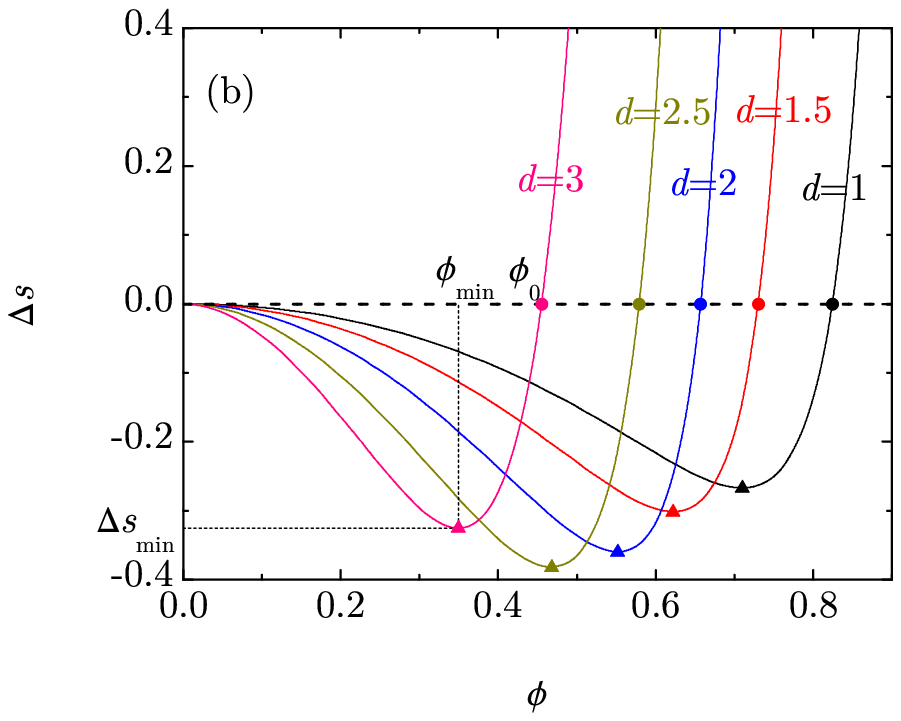}
\caption{(\textbf{a}) Plot of $s_\ex(\eeta)$  (solid lines) and $s_2(\eeta)$ (dashed lines) for dimensions $d=1$, $1.5$, $2$, $2.5$, and $3$. The circles indicate the points where $s_\ex(\eeta)$  and $s_2(\eeta)$ cross. (\textbf{b}) Plot of $\Delta s(\eeta)=s_\ex(\eeta)-s_2(\eeta)$  for  $d=1$, $1.5$, $2$, $2.5$, and $3$. The triangles indicate the location of the minima and the circles indicate the packing fractions $\eeta_0$ where $\Delta s=0$.
\label{fig1}}
\end{figure}

Figure \ref{fig1}a shows $s_\ex(\eeta)$   and $s_2(\eeta)$ as functions of the packing fraction for a few dimensions $1\leq d\leq 3$. In all the cases, both functions become more negative as the packing fraction increases. Moreover, at a common packing fraction $\eeta$, both $s_\ex(\eeta)$   and $s_2(\eeta)$ decrease as the dimensionality increases. This is an expected property in the conventional case of integer $d$ since, at a common $\eeta$, all the thermodynamic quantities depart more from their ideal-gas values with increasing $d$. Not surprisingly, this property is maintained in the case of noninteger $d$.

Figure \ref{fig1}a also shows that the pair entropy $s_2(\eeta)$ overestimates the excess entropy $s_\ex(\eeta)$ for packing fractions smaller than a certain value $\eeta_0$. This means that, if $\eeta<\eeta_0$, the cumulated effect of correlations involving three, four, five, \ldots particles produces a decrease of the entropy. The opposite situation occurs, however, if $\eeta>\eeta_0$. At the threshold point $\eeta=\eeta_0$ the cumulated effect of multiparticle correlations cancels and then only the pair correlations contribute to $s_\ex$.

The density dependence of the RMPE $\Delta s=s_\ex-s_2$ is shown in Fig.\ \ref{fig1}b for the same values of $d$ as in   Fig.\ \ref{fig1}a. The qualitative shape of $\Delta s(\eeta)$ is analogous for all $d$: $\Delta s$ starts with a zero value at $\eeta=0$, then decreases and reaches a minimum value $\Delta s_{\min}$ at a  certain packing fraction $\eeta_{\min}$, after which it grows very rapidly, crossing the zero value at the packing fraction $\eeta_0$.

\begin{figure}[H]
\centering
\includegraphics[width=.45\columnwidth]{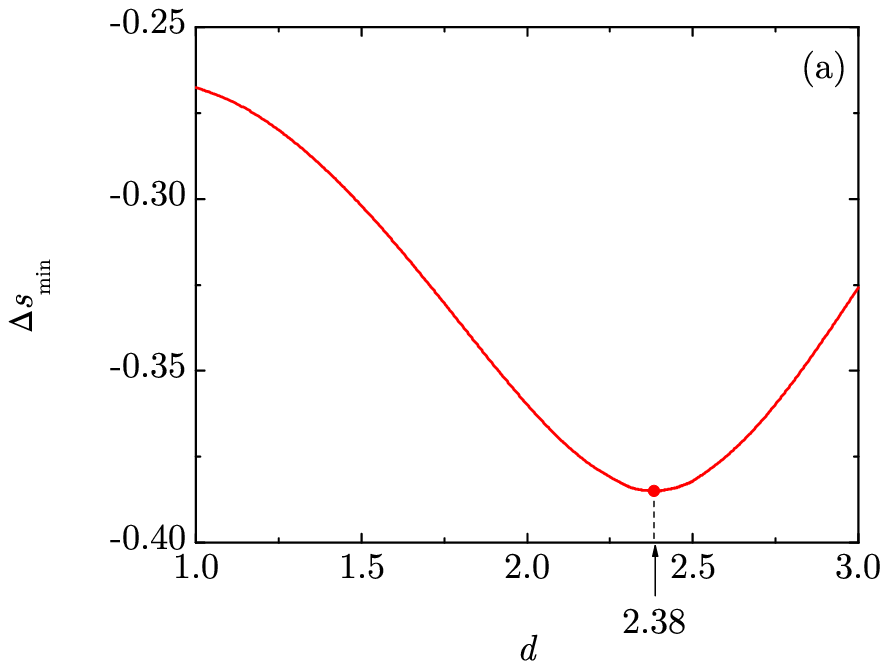}\hspace{1cm}\includegraphics[width=.45\columnwidth]{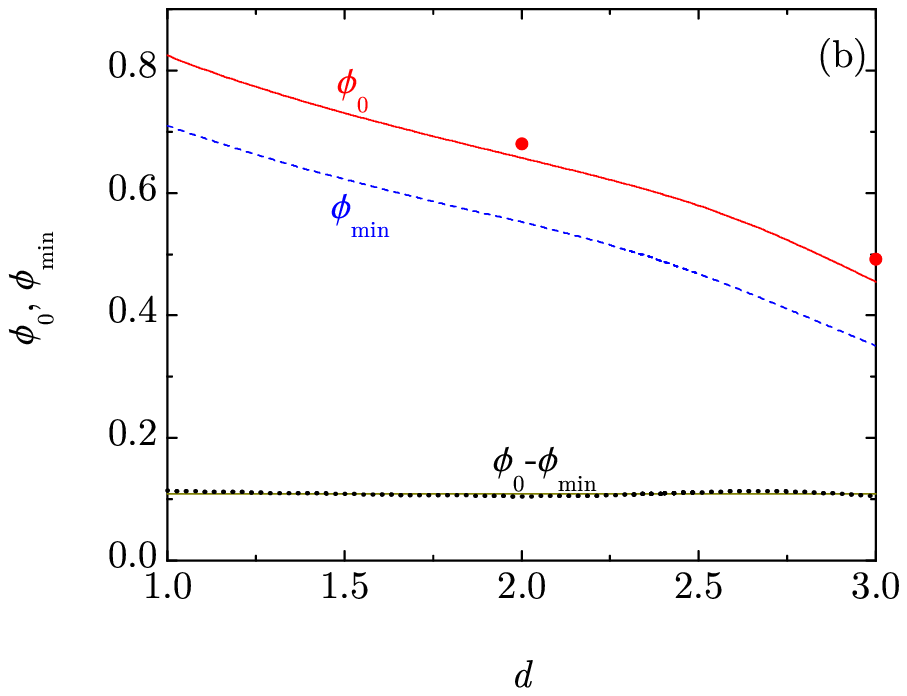}
\caption{(\textbf{a}) Plot of $\Delta s_{\min}$  as a function of $d$. The circle and the arrow indicate the location of the minimum at $d\simeq 2.383$. (\textbf{b}) Plot of $\eeta_0$  (solid line), $\eeta_{\min}$ (dashed line), and the difference $\eeta_0-\eeta_{\min}$ (dotted line) as functions of $d$. The horizontal solid line signals the value $\eeta_0-\eeta_{\min}=0.109$. The circles represent the values $\eeta=0.68$ at $d=2$ and $\eeta=0.49$ at $d=3$ corresponding to the fluid-hexatic \cite{AW62,TAAD17} and fluid-crystal \cite{AW57,FMSV12,RHS14} transitions, respectively.
\label{fig2}}
\end{figure}

The dimensionality dependence of the minimum value of the RMPE, $\Delta s_{\min}$, is displayed in Fig.\ \ref{fig2}a. Interestingly enough, as can also be observed in Fig.\ \ref{fig1}a, $\Delta s_{\min}$ presents  a nonmonotonic variation with $d$, having an absolute minimum $\Delta s_{\min}\simeq -0.385$ at $d\simeq 2.383$. At this noninteger dimensionality the pair entropy $s_2$ represents the largest overestimate of the excess entropy $s_\ex$. In contrast to $\Delta s_{\min}$, both $\eeta_0$ and $\eeta_{\min}$ decay monotonically with increasing $d$. This is clearly observed from Fig.\ \ref{fig2}b, where also the fluid-hexatic and the fluid-crystal transition points for disks and spheres, respectively, are shown. The proximity of those two points to the curve $\eeta_0$ provide support to the  zero-RMPE criterion, especially considering the approximate character of our simple theoretical approach. Thus, if a disorder-to-order transition phase is possible for fractal hard-core liquids, we expect that it is located near (possibly slightly above) the packing fraction $\eeta_0$.

\begin{figure}[H]
\centering
\includegraphics[width=.45\columnwidth]{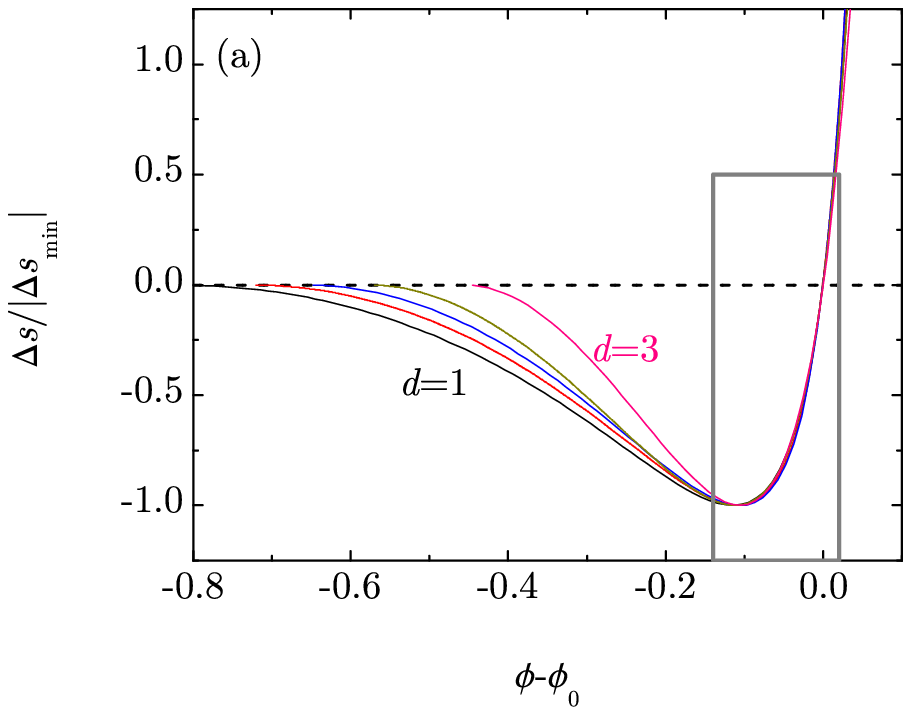}\hspace{1cm}\includegraphics[width=.45\columnwidth]{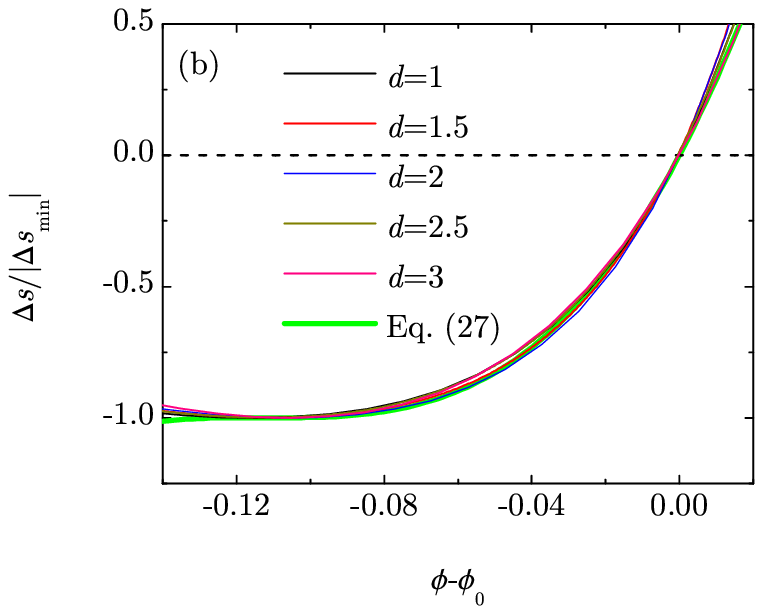}
\caption{(\textbf{a}) Plot of the scaled RMPE $\Delta s/|\Delta s_{\min}|$ as a function of the difference $\eeta-\eeta_0$ for dimensions $d=1$, $1.5$, $2$, $2.5$, and $3$.  (\textbf{b}) Magnification of the framed region of panel \textbf{a}. The light thick line represents the formula given by Eq.\ \eqref{fit}.
\label{fig3}}
\end{figure}

An interesting feature of Fig.\ \ref{fig2}b is that the difference $\eeta_0-\eeta_{\min}\simeq 0.109$ is hardly dependent on $d$. This suggests the possibility of a quasiuniversal behavior of the \emph{scaled} RMPE $\Delta s/|\Delta s_{\min}|$ in the neighborhood of $\eeta=\eeta_0$. To check this possibility, Fig.\ \ref{fig3}a shows $\Delta s/|\Delta s_{\min}|$ as a function of $\eeta-\eeta_0$ for the same dimensionalities as in Fig.\ \ref{fig1}. We can observe a relatively good collapse of the curves in the region $-0.14\lesssim \eeta-\eeta_0\lesssim 0.02$. A magnification of that region is shown in Fig.\ \ref{fig3}b. A simple fit can be obtained as follows. Let us define $X\equiv (\eeta-\eeta_0)/0.109$ and $Y(X)\equiv \Delta s(\eeta)/|\Delta s_{\min}|$. Then, a cubic function $Y(X)$ consistent with the conditions $Y(0)=0$, $Y(-1)=-1$, $Y'(-1)=0$, $Y''(-1)>0$ is $Y(X)=X\left[2+X+c (1+X)^2\right]$ with $c<1$. A good agreement is found with $0.8<c<1$ and we choose $c=0.9$. In summary, our proposed universal form is
\beq
\label{fit}
\frac{\Delta s(\eeta)}{|\Delta s_{\min}|}\simeq X\left[2+X+c (1+X)^2\right],\quad X\equiv \frac{\eeta-\eeta_0}{0.109},\quad c=0.9.
\eeq
It is also plotted in Fig.\ \ref{fig3}b, where we can see that it captures well the behavior for dimensions $1\leq d\leq 3$.

Before closing this section, it is convenient to add a comment. As said at the end of Sec.\ \ref{sec2},  the values of $\Delta s$ have been obtained from Eq.\ \eqref{Delta_s} by evaluating $\Ds_2$ from Eq.\ \eqref{8b} numerically.
Since in Eq.\ \eqref{7} we have followed the virial route, here we will refer to this method to obtain the function $\Delta s$ as the virial route and denote the resulting quantity as $\Delta s^\vir$. On the other hand, this method is not exactly equivalent to that obtained from Eq. \eqref{1} with $s_2$ evaluated numerically from Eq.\ \eqref{2a} by following the same procedure as described above for $\Ds_2$. This alternative method will be referred to as the compressibility route ($\Delta s^\comp$), since it is equivalent to evaluating the isothermal compressibility from Eq.\ \eqref{chiT}. Therefore, according to Eq.\ \eqref{Delta_s},
\beq
\Delta s^\vir-\Delta s^\comp=-\frac{1}{2}\left(\chi_T^\vir-\chi_T^\comp\right).
\eeq
We have checked that both methods (virial and compressibility) yield practically indistinguishable results. For instance, if $d=3$, $\eeta_0=0.4552$ in the virial route, while $\eeta_0=0.4547$ in the compressibility route. At $d=1$ and $d=2$ both methods yield, consistently, $\eeta_0=0.8246$ and $\eeta_0=0.6573$, respectively.
Note that the compressibility route to measure $\Delta s$ has still a virial ``relic'' in the contribution coming from the excess free energy, Eq.\ \eqref{6b}. A pure compressibility route would require the numerical evaluation of $\chi_T$ from Eq.\ \eqref{chiT} and then a double numerical integration, as evident from Eqs.\ \eqref{betaa_ex} and \eqref{4}. This procedure would complicate enormously the evaluation of $s_\ex$ without any significant gain in accuracy.
%
%%%%%%%%%%%%%%%%%%%%%%%%%%%%%%%%%%%%%%%%%%
\section{Conclusions}
\label{sec4}
In this article we have calculated the pair contribution and the cumulative contribution arising from correlations involving more than two particles to the excess entropy of hard spheres in fractional dimensions $1<d<3$. To this end, we have resorted to the analytical approximations for the equation of state and radial distribution function of the fluid previously set up by Santos and L\'opez de Haro \cite{SH16}. Over the fractional dimensionality range explored, the so-called ``residual multiparticle entropy'' (RMPE), obtained as the difference between the excess and pair entropies, shows a behavior utterly similar to that exhibited for integer 1, 2, and 3 dimensions. Hence, on a phenomenological continuity basis, we surmise that hard spheres undergo an ``ordering'' transition even in a space with fractional dimensions, which may well anticipate a proper thermodynamic fluid-to-solid phase transition.

We found that the packing fraction loci of minimum and vanishing RMPE show a monotonic decreasing behavior as a function of the dimensionality; this result is coherent with the magnification of excluded-volume effects produced by increasing spatial dimensionalities and, correspondingly, with a gradual shift of the ordering transition threshold to lower and lower packing fractions. However, it also turns out that the minimum value of the RMPE exhibits a non-monotonic behavior, attaining a minimum at the fractional dimensionality $d=2.383$. For this value of $d$ the relative entropic weight of more-than-two-particle correlations reaches, in the ``gas-like'' regime, its maximum absolute value.

Finally, the quasi-universal scaling of the RMPE over its minimum value in the neighborhood of the sign-crossover point suggests that the properties of the local ordering phenomenon should not sensitively depend on the spatial dimensionality.

%%%%%%%%%%%%%%%%%%%%%%%%%%%%%%%%%%%%%%%%%%

\vspace{6pt}
%%%%%%%%%%%%%%%%%%%%%%%%%%%%%%%%%%%%%%%%%%
%% optional
%\supplementary{The following are available online at \linksupplementary{s1}, Figure S1: title, Table S1: title, Video S1: title.}

% \supplementary{The following are available at \linksupplementary, Figure S1: title, Table S1: title, Video S1: title. A supporting video article is available at doi: link.}

%%%%%%%%%%%%%%%%%%%%%%%%%%%%%%%%%%%%%%%%%%
\authorcontributions{A.S. proposed the idea and performed the calculations;
F.S. and P.V.G. participated in the analysis and discussion of the results; the three authors worked on the revision and writing of the final manuscript.}

%%%%%%%%%%%%%%%%%%%%%%%%%%%%%%%%%%%%%%%%%%
\funding{A.S. acknowledges financial support from the Ministerio de Econom\'ia y Competitividad (Spain) through Grant No.\ FIS2016-76359-P and the Junta de Extremadura
(Spain) through Grant No.\ GR18079, both partially financed by Fondo Europeo de Desarrollo Regional funds.}

%%%%%%%%%%%%%%%%%%%%%%%%%%%%%%%%%%%%%%%%%%
\acknowledgments{A.S. is grateful to Dr.\ Roberto Trasarti-Battistoni for helpul discussions and for bringing Ref.\ \cite{LS91} to our attention.}

%%%%%%%%%%%%%%%%%%%%%%%%%%%%%%%%%%%%%%%%%%
\conflictsofinterest{The authors declare no conflict of interest. %The funding sponsors had no role in the design of the study; in the collection, analyses, or interpretation of data; in the writing of the manuscript, and in the decision to publish the results''.
}

%%%%%%%%%%%%%%%%%%%%%%%%%%%%%%%%%%%%%%%%%%
%% optional
\abbreviations{The following abbreviations are used in this manuscript:\\

\noindent
\begin{tabular}{@{}ll}
RMPE & Residual Multiparticle Entropy\\
MC  & Monte Carlo\\
PY & Percus--Yevick
\end{tabular}}

\reftitle{References}
%=====================================
% References, variant B: external bibliography
%=====================================
\externalbibliography{yes}
\bibliography{RMPE_fractal_HS}

\begin{thebibliography}{999}
\providecommand{\natexlab}[1]{#1}

\bibitem[Wong and Cao(1992)]{WC92}
Wong, P.z.; Cao, Q.z.
\newblock Correlation function and structure factor for a mass fractal bounded
  by a surface fractal.
\newblock {\em Phys. Rev. B} {\bf 1992}, {\em 45},~7627--7632,
\newblock
  doi:{\changeurlcolor{black}\href{https://doi.org/10.1103/PhysRevB.45.7627}{\detokenize{10.1103/PhysRevB.45.7627}}}.

\bibitem[Kurzidim \em{et~al.}(2009)Kurzidim, Coslovich, and Kahl]{KCK09}
Kurzidim, J.; Coslovich, D.; Kahl, G.
\newblock Single-Particle and Collective Slow Dynamics of Colloids in Porous
  Confinement.
\newblock {\em Phys. Rev. Lett.} {\bf 2009}, {\em 103}, 138303,
\newblock
  doi:{\changeurlcolor{black}\href{https://doi.org/10.1103/PhysRevLett.103.138303}{\detokenize{10.1103/PhysRevLett.103.138303}}}.

\bibitem[Kim \em{et~al.}(2011)Kim, Miyazaki, and Saito]{KMS11}
Kim, K.; Miyazaki, K.; Saito, S.
\newblock Slow dynamics, dynamic heterogeneities, and fragility of supercooled
  liquids confined in random media.
\newblock {\em J. Phys. Condens. Matter} {\bf 2011}, {\em 23},~234123,
\newblock
  doi:{\changeurlcolor{black}\href{https://doi.org/10.1088/0953-8984/23/23/234123}{\detokenize{10.1088/0953-8984/23/23/234123}}}.

\bibitem[Skinner \em{et~al.}(2013)Skinner, Schnyder, Aarts, Horbach, and
  Dullens]{SSAHD13}
Skinner, T.O.E.; Schnyder, S.K.; Aarts, D.G.A.L.; Horbach, J.; Dullens, R.P.A.
\newblock Localization Dynamics of Fluids in Random Confinement.
\newblock {\em Phys. Rev. Lett.} {\bf 2013}, {\em 111},~128301,
\newblock
  doi:{\changeurlcolor{black}\href{https://doi.org/10.1103/PhysRevLett.111.128301}{\detokenize{10.1103/PhysRevLett.111.128301}}}.

\bibitem[Heinen \em{et~al.}(2015)Heinen, Schnyder, Brady, and L\"owen]{HSBL15}
Heinen, M.; Schnyder, S.K.; Brady, J.F.; L\"owen, H.
\newblock Classical Liquids in Fractal Dimension.
\newblock {\em Phys. Rev. Lett.} {\bf 2015}, {\em 115},~097801,
\newblock
  doi:{\changeurlcolor{black}\href{https://doi.org/10.1103/PhysRevLett.115.097801}{\detokenize{10.1103/PhysRevLett.115.097801}}}.

\bibitem[Santos and { L\'opez de Haro}(2016)]{SH16}
Santos, A.; { L\'opez de Haro}, M.
\newblock Radial distribution function for hard spheres in fractal dimensions:
  A~heuristic approximation.
\newblock {\em Phys. Rev. E} {\bf 2016}, {\em 93},~062126,
\newblock
  doi:{\changeurlcolor{black}\href{https://doi.org/10.1103/PhysRevE.93.062126}{\detokenize{10.1103/PhysRevE.93.062126}}}.

\bibitem[Nettleton and Green(1958)]{Nettleton1958}
Nettleton, R.E.; Green, M.S.
\newblock Expression in Terms of Molecular Distribution Functions for the
  Entropy Density in an Infinite System.
\newblock {\em J. Chem. Phys.} {\bf 1958}, {\em 29},~1365--1370,
\newblock
  doi:{\changeurlcolor{black}\href{https://doi.org/10.1063/1.1744724}{\detokenize{10.1063/1.1744724}}}.

\bibitem[Baranyai and Evans(1989)]{Baranyai1989}
Baranyai, A.; Evans, D.J.
\newblock Direct entropy calculation from computer simulation of liquids.
\newblock {\em Phys. Rev. A} {\bf 1989}, {\em 40},~3817--3822,
\newblock
  doi:{\changeurlcolor{black}\href{https://doi.org/10.1103/PhysRevA.40.3817}{\detokenize{10.1103/PhysRevA.40.3817}}}.

\bibitem[Giaquinta and Giunta(1992)]{GG92}
Giaquinta, P.V.; Giunta, G.
\newblock About entropy and correlations in a fluid of hard spheres.
\newblock {\em Phys. A} {\bf 1992}, {\em 187},~145--158,
\newblock
  doi:{\changeurlcolor{black}\href{https://doi.org/10.1016/0378-4371(92)90415-M}{\detokenize{10.1016/0378-4371(92)90415-M}}}.

\bibitem[Giaquinta(2008)]{G08}
Giaquinta, P.V.
\newblock Entropy and Ordering of Hard Rods in One Dimension.
\newblock {\em Entropy} {\bf 2008}, {\em 10},~248--260,
\newblock
  doi:{\changeurlcolor{black}\href{https://doi.org/10.3390/e10030248}{\detokenize{10.3390/e10030248}}}.

\bibitem[Krekelberg \em{et~al.}(2008)Krekelberg, Shen, Errington, and
  Truskett]{KSET08}
Krekelberg, W.P.; Shen, V.K.; Errington, J.R.; Truskett, T.M.
\newblock Residual multiparticle entropy does not generally change sign near
  freezing.
\newblock {\em J. Chem. Phys.} {\bf 2008}, {\em 128},~161101,
\newblock
  doi:{\changeurlcolor{black}\href{https://doi.org/10.1063/1.2916697}{\detokenize{10.1063/1.2916697}}}.

\bibitem[Krekelberg \em{et~al.}(2009)Krekelberg, Shen, Errington, and
  Truskett]{KSET09}
Krekelberg, W.P.; Shen, V.K.; Errington, J.R.; Truskett, T.M.
\newblock Response to ``Comment on {`}Residual multiparticle entropy does not
  generally change sign near {freezing}'† [J. Chem. Phys. 130, 037101
  (2009)].
\newblock {\em J. Chem. Phys.} {\bf 2009}, {\em 130},~037102,
\newblock
  doi:{\changeurlcolor{black}\href{https://doi.org/10.1063/1.3058798}{\detokenize{10.1063/1.3058798}}}.

\bibitem[Giaquinta(2009)]{G09}
Giaquinta, P.V.
\newblock Comment on ``Residual multiparticle entropy does not generally change
  sign near freezing" {[J. Chem. Phys. 128, 161101 (2008)]}.
\newblock {\em J. Chem. Phys.} {\bf 2009}, {\em 130},~037101,
\newblock
  doi:{\changeurlcolor{black}\href{https://doi.org/10.1063/1.3058794}{\detokenize{10.1063/1.3058794}}}.

\bibitem[Saija \em{et~al.}(1998)Saija, Pastore, and Giaquinta]{Saija1998}
Saija, F.; Pastore, G.; Giaquinta, P.V.
\newblock Entropy and Fluid-Fluid Separation in Nonadditive Hard-Sphere
  Mixtures.
\newblock {\em J. Phys. Chem. B} {\bf 1998}, {\em 102},~10368--10371,
\newblock
  doi:{\changeurlcolor{black}\href{https://doi.org/10.1021/jp982202b}{\detokenize{10.1021/jp982202b}}}.

\bibitem[Costa \em{et~al.}(2002)Costa, Micali, Saija, and Giaquinta]{Costa2002}
Costa, D.; Micali, F.; Saija, F.; Giaquinta, P.V.
\newblock Entropy and Correlations in a Fluid of Hard Spherocylinders: The
  Onset of Nematic and Smectic Order.
\newblock {\em J. Phys. Chem. B} {\bf 2002}, {\em 106},~12297--12306,
\newblock
  doi:{\changeurlcolor{black}\href{https://doi.org/10.1021/jp0259317}{\detokenize{10.1021/jp0259317}}}.

\bibitem[Saija \em{et~al.}(2003)Saija, Saitta, and Giaquinta]{Saija2003}
Saija, F.; Saitta, A.M.; Giaquinta, P.V.
\newblock Statistical entropy and density maximum anomaly in liquid water.
\newblock {\em J.~Chem. Phys.} {\bf 2003}, {\em 119},~3587--3589,
\newblock
  doi:{\changeurlcolor{black}\href{https://doi.org/10.1063/1.1598431}{\detokenize{10.1063/1.1598431}}}.

\bibitem[Banerjee \em{et~al.}(2017)Banerjee, Nandi, Sastry, and
  Bhattacharyya]{BNSMB2017}
Banerjee, A.; Nandi, M.K.; Sastry, S.; Bhattacharyya, S.M.
\newblock Determination of onset temperature from the entropy for fragile to
  strong liquids.
\newblock {\em J. Chem. Phys.} {\bf 2017}, {\em 147},~024504,
\newblock
  doi:{\changeurlcolor{black}\href{https://doi.org/10.1063/1.4991848}{\detokenize{10.1063/1.4991848}}}.

\bibitem[Santos(2016)]{S16}
Santos, A.
\newblock A Concise Course on the Theory of Classical Liquids. Basics and
  Selected Topics. In \emph{Lecture Notes in Physics}; Springer: New
  York, NY, USA,  2016; Volume 923.

\bibitem[Lemson and Sanders(1991)]{LS91}
Lemson, G.; Sanders, R.H.
\newblock On the use of the conditional density as a description of galaxy
  clustering.
\newblock {\em Mon.~Not. R. Astron. Soc.} {\bf 1991}, {\em 252},~319--328,
\newblock
  doi:{\changeurlcolor{black}\href{https://doi.org/10.1093/mnras/252.3.319}{\detokenize{10.1093/mnras/252.3.319}}}.

\bibitem[Barker and Henderson(1976)]{BH76}
Barker, J.A.; Henderson, D.
\newblock What is ``liquid"? {Understanding} the states of matter.
\newblock {\em Rev. Mod. Phys.} {\bf 1976}, {\em 48},~587--671,
\newblock
  doi:{\changeurlcolor{black}\href{https://doi.org/10.1103/RevModPhys.48.587}{\detokenize{10.1103/RevModPhys.48.587}}}.

\bibitem[Percus and Yevick(1958)]{PY58}
Percus, J.K.; Yevick, G.J.
\newblock Analysis of Classical Statistical Mechanics by Means of Collective
  Coordinates.
\newblock {\em Phys. Rev.} {\bf 1958}, {\em 110},~1--13,
\newblock
  doi:{\changeurlcolor{black}\href{https://doi.org/10.1103/PhysRev.110.1}{\detokenize{10.1103/PhysRev.110.1}}}.

\bibitem[Henderson(1975)]{H75}
Henderson, D.
\newblock A simple equation of state for hard discs.
\newblock {\em Mol. Phys.} {\bf 1975}, {\em 30}, 971--972,
\newblock
  doi:{\changeurlcolor{black}\href{https://doi.org/10.1080/00268977500102511}{\detokenize{10.1080/00268977 500102511}}}.

\bibitem[Wertheim(1963)]{W63}
Wertheim, M.S.
\newblock Exact solution of the {Percus-Yevick} integral equation for hard
  spheres.
\newblock {\em Phys. Rev. Lett.} {\bf 1963}, {\em 10},~321--323,
\newblock
  doi:{\changeurlcolor{black}\href{https://doi.org/10.1103/PhysRevLett.10.321.}{\detokenize{10.1103/PhysRevLett.10.321.}}}

\bibitem[Thiele(1963)]{T63}
Thiele, E.
\newblock Equation of state for hard spheres.
\newblock {\em J. Chem. Phys.} {\bf 1963}, {\em 39},~474--479,
\newblock
  doi:{\changeurlcolor{black}\href{https://doi.org/10.1063/1.1734272}{\detokenize{10.1063/1.1734272}}}.

\bibitem[Alder and Wainwright(1962)]{AW62}
Alder, B.J.; Wainwright, T.E.
\newblock Phase Transition in Elastic Disks.
\newblock {\em Phys. Rev.} {\bf 1962}, {\em 127}, 359--361,
\newblock
  doi:{\changeurlcolor{black}\href{https://doi.org/10.1103/PhysRev.127.359}{\detokenize{10.1103/ PhysRev.127.359}}}.

\bibitem[Thorneywork \em{et~al.}(2017)Thorneywork, Abbott, Aarts, and
  Dullens]{TAAD17}
Thorneywork, A.L.; Abbott, J.L.; Aarts, D.G.A.L.; Dullens, R.P.A.
\newblock Two-Dimensional Melting of Colloidal Hard Spheres.
\newblock {\em Phys. Rev. Lett.} {\bf 2017}, {\em 118},~158001,
\newblock
  doi:{\changeurlcolor{black}\href{https://doi.org/10.1103/PhysRevLett.118.158001}{\detokenize{10.1103/PhysRevLett.118.158001}}}.

\bibitem[Alder and Wainwright(1957)]{AW57}
Alder, B.J.; Wainwright, T.E.
\newblock Phase Transition for a Hard Sphere System.
\newblock {\em J. Chem. Phys.} {\bf 1957}, {\em 27},~1208--1209,
\newblock
  doi:{\changeurlcolor{black}\href{https://doi.org/10.1063/1.1743957}{\detokenize{10.1063/1.1743957}}}.

\bibitem[Fern\'andez \em{et~al.}(2012)Fern\'andez, Mart\'in-Mayor, Seoane, and
  Verrocchio]{FMSV12}
Fern\'andez, L.A.; Mart\'in-Mayor, V.; Seoane, B.; Verrocchio, P.
\newblock Equilibrium Fluid-Solid Coexistence of Hard Spheres.
\newblock {\em Phys. Rev. Lett.} {\bf 2012}, {\em 108},~{165701},
\newblock
  doi:{\changeurlcolor{black}\href{https://doi.org/10.1103/PhysRevLett.108.165701}{\detokenize{10.1103/PhysRevLett.108.165701}}}.

\bibitem[Robles \em{et~al.}(2014)Robles, {L\'opez de Haro}, and Santos]{RHS14}
Robles, M.; {L\'opez de Haro}, M.; Santos, A.
\newblock Note: Equation of state and the freezing point in the hard-sphere
  model.
\newblock {\em J. Chem. Phys.} {\bf 2014}, {\em 140},~136101,
\newblock
  doi:{\changeurlcolor{black}\href{https://doi.org/10.1063/1.4870524}{\detokenize{10.1063/1.4870524}}}.

\bibitem[Vannimenus \em{et~al.}(1984)Vannimenus, Nadal, and Martin]{VNM84}
Vannimenus, J.; Nadal, J.P.; Martin, H.
\newblock On the spreading dimension of percolation and directed percolation clusters.
\newblock {\em J. Phys. A: Math. Gen.} {\bf 1984}, {\em 17},~L351--L356,
\newblock
  doi:{\changeurlcolor{black}\href{https://doi.org/10.1088/0305-4470/17/6/008}{\detokenize{10.1088/0305-4470/17/6/008}}}.

\bibitem[ben-Avraham and Havlin(2000)]{AH00}
ben-Avraham, D.; Havlin, S.
\newblock {\em {Diffusion and Reactions in Fractal and Disordered Systems}}; Cambridge University Press: Cambridge, UK, 2016.

\end{thebibliography}

%%%%%%%%%%%%%%%%%%%%%%%%%%%%%%%%%%%%%%%%%%
%% optional
%\sampleavailability{Samples of the compounds ...... are available from the authors.}

%% for journal Sci
%\reviewreports{\\
%Reviewer 1 comments and authors’ response\\
%Reviewer 2 comments and authors’ response\\
%Reviewer 3 comments and authors’ response
%}

%%%%%%%%%%%%%%%%%%%%%%%%%%%%%%%%%%%%%%%%%%
\end{document}